\documentclass[acmsmall]{acmart}

\AtBeginDocument{%
  }

\setcopyright{acmlicensed}
\copyrightyear{2018}
\acmYear{2018}
\acmDOI{XXXXXXX.XXXXXXX}

\acmConference[Conference acronym 'XX]{Make sure to enter the correct
  conference title from your rights confirmation emai}{June 03--05,
  2018}{Woodstock, NY}

\acmISBN{978-1-4503-XXXX-X/18/06}
\usepackage{tikz}
\usetikzlibrary{shapes.geometric, arrows}
\usepackage{soul}
\usepackage[table]{xcolor}
\usepackage{tabularx}
\usepackage{array}

\sethlcolor{yellow}

\usepackage{booktabs}         
\usepackage{colortbl}         

\definecolor{RomanceCol}{HTML}{DFF5EE} 
\definecolor{HeaderGray}{gray}{0.95}

\begin{document}
\copyrightyear{2026}
\acmYear{2026}
\setcopyright{cc}
\setcctype{by}
\acmConference[CHI '26]{Proceedings of the 2026 CHI Conference on Human Factors in Computing Systems}{April 13--17, 2026}{Barcelona, Spain}
\acmBooktitle{Proceedings of the 2026 CHI Conference on Human Factors in Computing Systems (CHI '26), April 13--17, 2026, Barcelona, Spain}
\acmPrice{}
\acmDOI{10.1145/3772318.3791198}
\acmISBN{979-8-4007-2278-3/2026/04}


\title[Online Dating Romance Scams Targeting Minors in Iran]{“My Parents’ Expectations Were Overwhelming”: Online Dating Romance Scams Targeting Minors in Iran Through Exploitation of Parental Pressure}







\author{Sima Amirkhani}
\orcid{0009-0009-8836-8291}
\email{Sima.Amirkhani@h-brs.de}
\affiliation{%
\institution{Human-Computer Interaction, University of Siegen}
\city{Siegen}
\country{Germany}
}
\author{Mahla Alizadeh}
\orcid{0000-0002-5365-4695}
\affiliation{%
\institution{Human-Computer Interaction, University of Siegen}
\city{Siegen}
\country{Germany}}
\email{fatemeh.alizadeh@uni-siegen.de}

\author{Dave Randall}
 \orcid{0000-0001-8613-3477}
\affiliation{%
\institution{Human-Computer Interaction, University of Siegen}
\city{Siegen}
\country{Germany}}

\author{Gunnar Stevens}
 \orcid{0000-0002-7785-5061}
\affiliation{%
\institution{Human computer Interaction, Hochschule Bonn-Rhein-Sieg}
\city{Sankt Augustin}
\country{Germany}}

\author{Douglas Zytko}
\orcid{0000-0002-6854-5336}
\affiliation{%
\institution{University of Michigan-Flint}
\city{Flint}
\state{Michigan}
\country{United States}
}

\renewcommand{\shortauthors}{Sima Amirkhani et al.}

\begin{abstract}

Minors are at risk of myriad harms online, yet online dating romance scams are seldom considered one of them. While research of romance scams in Western countries finds victims to predominantly be middle-age, it is unknown if minors in geographic regions with cultural norms around teenage marriage are uniquely susceptible to online dating romance scams. We present an interview study with 16 victims of online dating romance scams in Iran who were minors when scammed. Findings show that, with westernized dating apps banned in Iran, scammers find teenage victims through messaging platforms tethered to local neighborhoods, offering relief for parental pressures around finding a marital partner and academic performance. Using threats, lies, and exploitation of emotional attachment lacking from their families, scammers pressured minors into financial and sexual favors. The study demonstrates how local cultural context should be foregrounded in future research on, and solutions for, technology-mediated harm against minors. \textit{\textbf{Content Warning:} This paper discusses sexual abuse.}

\end{abstract}



\received{20 February 2007}
\received[revised]{12 March 2009}
\received[accepted]{5 June 2009}

\maketitle


\section{Introduction}

\textit{"In our region, when a girl turns 13, the family starts pressuring her to marry. If no one proposes, it means she’s not lovable. It becomes a competition. Girls rush to marry so they can say they won someone’s love. So when someone online told me he loved me and wanted to marry me, I was already in that mindset." } - Ziba (P13)

Since the early 2000s, romance scams have been a pervasive threat in the online dating landscape. However, they are not
typically associated with minors \cite{muniz2023victims} or with the array of technology-mediated harms against children typically studied in HCI \cite{chang2025generating,agha2025systematic, namvarpour2025romance,wisniewski2014adolescent, livingstone2009kids} such as grooming \cite{greene2020experiences, razi2020let}, sexual solicitation \cite{dev2022ignoring,gamez2018persuasion, razi2020let}, harassment \cite{iivari2021chi}, and cyberbullying \cite{ashktorab2016designing}. Online dating romance scams involve the use of a fake identity, through a dating app or social platform, to manipulate a victim into a false romantic relationship to gain items of value \cite{garrett2014exploring}; often through financial fraud \cite{whitty2015anatomy} as well as sextortion \cite{cross2023pay}. 
Victims experience profound emotional and psychological harm \cite{wang2022online, yosiandra2024unveiling,wang2024don}.
 
Prior work often portrays online dating romance scam victims as middle-aged, well-educated women 
in Western countries \cite{wiederhold2024digital, buchanan2014online, edwards2018geography}. 
This Western-centric profile risks overlooking other types of victims, such as minors in 
non-Western cultures where expectations around dating and marriage may be imposed during adolescence.
For example, in Iran, the legal minimum age for marriage is 13, or as early as 9 with parental approval, especially in rural areas \cite{asghari2019early}. 
If, and how, minors in such contexts experience technology-mediated romance scams remains unknown, representing a gap in both romance scam research and HCI scholarship on technology-mediated harms against children \cite{shaari2019online, alsoubai2022friends, chen1997authoritative}.

To address this gap, we present an interview study with 16 victims of online dating romance scams in Iran—10 from rural areas and 6 from urban areas—who were minors at the time of the scam. The research questions driving our study were: \textit{\textbf{RQ1. }What strategies are used by online dating romance scammers who target minors in Iran? \textbf{RQ2.} What role does Iranian dating culture play in minors’ susceptibility to online dating romance scams? \textbf{RQ3.} How do such scams end, and what are the post-scam impacts on minors who were victimized?}

 Findings show that romance scams against minors in Iran exploit local cultural pressures imposed or reinforced by the victims’ parents. With westernized dating apps being banned in Iran, scammers instead found their victims through Telegram and other messaging platforms with group chats tethered to the victim’s neighborhood. Through prolonged messaging interactions, scammers offered relief from cultural pressures imposed or reinforced by the victim’s parents, specifically the expectation of finding a marital partner while a teenager, pressure to perform well in school, and a more general lack of personal agency due to extreme parental oversight. Through a combination of threats, lies, and exploitation of emotional attachment that victims otherwise lacked in their families, scammers pressured victims into financial assistance and sexual favors. Outcomes resulted in severe mental health impacts on victims, resulting in multiple suicide attempts for one participant, in part due to negative judgment from family members.

The study makes three key contributions. First, we frame online dating romance scams targeting minors as a distinct form of technology-mediated harm against children in an understudied context. Second, we show how local cultural dynamics and scammers’ strategies for manipulating victims are tightly interwoven, demonstrating that HCI research into harm against minors must centre on cultural context. Finally, we offer design implications for digital resilience, highlighting how platforms and policymakers can better protect minors in socio-technical ecosystems.


\section{Related Work }
Research on online risks for young people spans diverse areas, from sexual grooming and cyberbullying to scams and fraud, yet important conceptual and contextual gaps remain.
To situate our work, we review related literature along three dimensions: first, exploring the literature to understand online dating romance scams (Section 2.1); second, minors’ online risks and vulnerabilities, including parenting influences (Section 2.2); and finally, the cultural context of romance in Iran, which shapes minors’ online interactions and their exposure to scams (Section 2.3).


\subsection{Online Dating Romance Scams}

Online dating romance scams are a form of fraud in which scammers construct fake identities on dating apps or other social platforms to deceive victims into believing they are in a genuine romantic relationship \cite{buchanan2014online, whitty2012online}. The intent of such scams is to extract \textit{"items of value"} \cite{garrett2014exploring}, such as money \cite{whitty2013scammers, rege2009s, cross2018denying} and sexual content \cite{cross2023pay, cross2024if}. 

Online dating romance scams generally unfold in several stages \cite{whitty2013scammers}, starting with the creation of a fabricated and idealized persona of a romantic partner \cite{kam2024drawing}. Once contact is made with the victim, typically through private messaging, a grooming phase begins \cite{shaari2019online}. Scammers leverage social engineering (SE), described as hacking people rather than systems to manipulate victims through exploitation of emotions and trust \cite{bharne2023enhanced, mesch2018low, alizadeh2023catch}. They often target individuals based on their online behavior and personal disclosures, selecting victims who exhibit traits like low self-esteem or high emotional vulnerability \cite{mesch2018low, wang2022online}. Scammers employ a range of manipulative techniques to foster emotional connections, frequently using romantic imagery and emotional appeals to deepen the relationship \cite{chuang2021romance, whitty2013scammers}.

Once grooming has fostered a strong emotional attachment from the victim, the scammer will leverage the supposed relationship to ask for or demand their desired item(s) of value. These are often financial in nature \cite{kopp2015role, wang2023persuasive}, typically through false pretenses of a medical emergency or other urgent situation that encourages the victim to act quickly \cite{whitty2013scammers}. This phase is also known to involve sextortion, through which the scammer solicits sexual content as its own item of value or as a means towards further financial gain by threatening to release the sexual content publicly \cite{wang2023persuasive, cross2023pay, cross2024if}.

The emotional and psychological costs to victims are substantial. Many experience what has been described as a “double hit”—the loss of both money and the perceived romantic relationship \cite{whitty2019can}. Victims often report long-lasting emotional distress, social shame, and isolation \cite{chen2022trauma, tan2013preying, whitty2016online}. Despite the profound and varied impacts on victims, existing research has disproportionately centred on adult victims—particularly middle-aged women in Western countries \cite{rege2009s, whitty2012online}. This narrow demographic focus has led to an under-theorization of how scammers adapt their tactics to different victim profiles. 
Strategies effective on a divorced woman in the U.S. (e.g., urgent requests to protect a “shared future”) \cite{whittle2013review} may differ dramatically from tactics used on a minor in Iran. Minors are a particularly vulnerable yet understudied group in online dating\cite{muniz2017online, muniz2023victims}. Their developmental and social contexts differ from adults, requiring focused analysis \cite{coluccia2020online, wang2024understanding, birnholtz2020layers, erikson1968identity}.

\subsection{Minors’ Online Risks and Vulnerabilities}
Minors today navigate an increasingly complex digital landscape, where opportunities for learning, socialization, and self-expression coexist with significant risks. Research in Human-Computer Interaction (HCI) and online safety has underscored that these risks are not uniform \cite{oguine2025online, oguine2025teens, amirkhani2025privacy}; they intersect with developmental, social, and cultural factors that shape how young people experience and respond to online threats \cite{oguine2025online, oguine2025teens, hartikainen2021safe, alluhidan2025unfiltered,agha2023strike}.  
Below, we review how these contextual factors interact with specific forms of harm, from sexual exploitation to cyberbullying, and highlight overlooked areas such as minors’ involvement in romance scams.

\subsubsection{Technology-Mediated Harms Targeting Minors}

Minors are vulnerable online to a broad spectrum of harms including grooming, sexual solicitation, sextortion, trafficking, harassment, cyberbullying, doxing, and exposure to harmful media content \cite{hamm2015prevalence, giumetti2022cyberbullying, chen2019doxing, alsoubai2022friends}. Research shows that individuals aged 13 to 18 spend on average over six hours daily online, with many active during late-night hours, increasing their vulnerabilities \cite{stoilova2021investigating, maloney2020virtual,alsoubai2024profiling, latzer2015disordered}. Theoretical frameworks such as Lifestyle and Routine Activity Theory (LRAT) suggest that increased online activity raises the likelihood of encountering offenders \cite{lee2022phishing, suh2020lifestyle}. Developmental vulnerabilities—including impulsivity, emotional naivety, and secrecy in relationships—further exacerbate  susceptibility to manipulation and exploitation \cite{wisniewski2014adolescent,stoilova2021investigating, maloney2020virtual, huh2023help}.

HCI and social computing research has primarily focused on grooming, solicitation, and sextortion as major online threats to minors, often framed within Western cultural and regulatory contexts \cite{ortega2022epidemiology, hsieh2023understanding, hornor2022online, hsieh2024correction, giumetti2022cyberbullying,hamm2015prevalence, kopczewski2018cyber} 
However, minors in diverse cultural settings may 
face distinct and understudied risks. Cultural taboos around sexuality and dating frequently hinder reporting and help-seeking, creating a dangerous silence exploited by offenders \cite{alsoubai2022friends, amirkhani2025society}.

Beyond sexual harms, minors also contend with cyberbullying, exposure to violent or sexualized media, body dissatisfaction, and psychological distress, underscoring the need for holistic approaches to safety \cite{hall2009fearnot, hamm2015prevalence, hinduja2010bullying,hinduja2008cyberbullying, wisniewski2014adolescent, orbach1998fat}.

\subsubsection{The Role of Parenting 
in Minors’ Online Vulnerability.}

Because minors are not adults, their parents play an important role in their online safety and risk exposure \cite{maloney2020virtual}. 
For example, tech-savvy parenting may foster safer online engagement, while overly restrictive or indulgent parenting styles may inadvertently increase online risks \cite{wisniewski2014adolescent, zweig2013rate}.
Ammari et al. \cite{ammari2015managing} explore how parents decide what personal information about their children to share on social networking sites, further highlighting parental roles in managing children’s safety. Alsoubai et al. \cite{alsoubai2022friends} emphasize the need for risk prevention strategies to protect youth from online sexual violence. Technology—when thoughtfully designed—can play an important role in augmenting parental efforts towards children's safety \cite{hall2009fearnot}, but sometimes technology can be obstructive of such efforts as well through privacy-invasive design choices \cite{ekambaranathan2021money, ekambaranathan2023can}.

Parenting styles can be an important influence on online risk through their shaping of adolescents’ psychological development and online behaviors. Positive family communication has shown strong protective effects in fostering minors’ online safety \cite{buelga2015family}. By contrast, authoritarian parenting—characterized by strict rules, low warmth, and high control—has been associated with negative psychological outcomes including low self-esteem, distrust, and social challenges \cite{simons2007linking}. This parenting style may influence minors’ susceptibility to online deception, cyberaggression, and victimization, especially within romantic digital interactions \cite{duran2015cyberbullying, rodriguez2018study}. While parental mediation of internet-use can reduce cyber risks, excessive restrictions may be ineffective or even counterproductive during adolescence \cite{lwin2008protecting, ekambaranathan2021money}. Evidence suggests that strict parenting may drive some adolescents to engage in covert or “stealthy” relationships online, increasing exposure to risks like dating violence and online scams \cite{muniz2023victims, muniz2019parental}. In light of this burgeoning insight on parenting styles, the influence of authoritarian parenting on minors’ romantic behaviors and decision-making in digital spaces necessitates further research.

\subsubsection{The Overlooked Intersection: Beyond Dating Violence, Minors as Victims of Online Dating Romance Scams}

Rivaz et~al., through their focus on adolescents as victims of \textit{offline} dating violence explicitly call for deeper inquiry into \textit{online} dating with under 18s: \textit{“we must bear in mind that the present study on dating violence victimization in adolescence did not include online victimization, an emerging problem that materializes in various ways"} (p. 133) and \textit{"Research on this type of victimization [dating victimization] and adjustment problems related to online environments is scarce.”} (p. 134)
\cite{muniz2023victims}. 
While research on minors’ online risks has predominantly centered on grooming and sexual solicitation \cite{maloney2020virtual,ammari2015managing,ekambaranathan2021money}, the body of work on romance scams largely examines only adult victims \cite{whitty2023drug,buchanan2014online,rege2009s}. In the adolescence literature, minors are frequently framed as aggressors in dating violence or as deviant/risk-taking users (e.g., trolling), rather than as victims of sustained, fraud-based romantic exploitation. This emphasis on perpetration and problematic behavior obscures how minors can be systematically targeted through long-term emotional manipulation, trust-building, and subsequent sexual or financial exploitation—especially in cultural contexts where dating norms and family expectations differ substantially \cite{amirkhani2025society}. Our work foregrounds an under-explored space: minors as victims of online dating romance scams, rather than as risk-takers or perpetrators.

\subsection{Iran: Cultural Contexts of Romance}

Prior work in HCI emphasizes that dating and romantic interaction online are deeply shaped by cultural contexts \cite{al2017against, al2021saudi}. In many non-Western settings, romantic discourse intersects with strict moral codes, family involvement, and legal constraints. For example, in India and parts of the Middle East, arranged marriages remain common, and dating outside family-sanctioned channels often carries stigma \cite{pasupathi2002arranged, parkin2021arranged,roudsari2013socio}. In conservative Asian societies, online dating may be perceived as morally questionable, leading to secretive or pseudonymous interactions \cite{liu2022cruel, low2022online, mehrotra2016south}. Similarly, in Iran, dating is frequently covert due to the criminalization of premarital relationships \cite{farahani2020adolescents, rahbari2016premarital}, which drives many to seek romantic connections through internet-based platforms \cite{mohammadi2006reproductive, amirkhani2023taking}.

This section examines the Iranian context to illustrate how cultural taboos, legal prohibitions, and family hierarchies intensify the vulnerability of minors to online romance scams—while simultaneously limiting avenues for disclosure or prevention.

\subsubsection{Cultural and Legal Structure Around Pre-Marital Romance}

Iran, with its 3,000-year-old cultural heritage, underwent a dramatic legal and social shift after the 1979 Islamic Revolution. Reforms that once aligned with Western family law were reversed, and conservative norms reinstated. The minimum legal marriage age dropped to 9 for girls and 15 for boys \cite{asghari2019early}, and laws once again permitted polygamy and reinforced male authority in divorce proceedings \cite{havenayereh}.

Romantic and sexual relationships outside marriage are not only socially stigmatized but legally criminalized \cite{farahani2020adolescents, rahbari2016premarital}. The Islamic Penal Code strictly regulates gender interactions, criminalizing physical and even verbal intimacy between unmarried men and women. Offenses such as \textit{Mozajeeh} (lying together) and \textit{Taqbil} (kissing) can incur up to 99 lashes under Article 637 \cite{boostani2022iranian, bahramitash2006myths}. More severe infractions, like \textit{Zena} (sexual intercourse outside marriage), carry punishments from 100 lashes to execution, as outlined in Articles 221, 224, and 225 \cite{shaheed2018outlier, maftei2010sanctions, zar2008case}.

Within this legal framework, marriage is regarded as the sole legitimate context for intimacy and reproduction. Childbearing outside marriage is forbidden, and early marriage—especially for girls—remains encouraged in many communities \cite{keshavarz2018desire, azadarmaki2006families}. While recent decades show a rising average marriage age among urban youth \cite{keshavarz2018desire}, traditional expectations persist. Girls often face pressure to marry young, sometimes internalizing societal ideals of “maturity” and “worthiness” at an early age \cite{hosseini2022reasons, mardi2018perceptions}.

Rigid norms make formal marriage inaccessible for many young people, leading them to pursue romantic ties secretly—seeking emotional intimacy or “love marriages” while avoiding legal and familial repercussions \cite{farahani2015meta, amirkhani2025society}. Urban centers such as Tehran show visible shifts, influenced by global media and digital technologies that introduce alternative ideals of dating and autonomy \cite{emami2021girlboy}. Yet these shifts coexist with strong resistance, creating a cultural divide between urban and rural regions and forcing youth to negotiate between evolving desires and entrenched constraints.

Despite these evolving dynamics, secrecy remains a dominant feature of young people’s romantic lives \cite{atari2020foundations, amirkhani2025society}. Social media platforms such as Instagram and Telegram allow Iranian youth to bypass familial and societal monitoring \cite{durrant2011secret, yadegarfard2019iranian}. Moreover, as mentioned by Joorabchi \cite{joorabchi2025examining} the features of Instagram further support the development of social connections and romantic interactions, especially among adolescents and young adults.
Meanwhile, parental expectations—emphasizing obedience, academic achievement, and moral purity—limit autonomy and discourage open dialogue about relationships \cite{salimi2005association, assadi2011beliefs, naeemi2024problematic}. The internet, therefore, becomes a crucial space for romantic connection, while simultaneously amplifying the tension between public and private selves \cite{yadegarfard2019iranian}. This often produces “split identities,” as traditional norms collide with global influences, shaping complex negotiations of authenticity, desire, and conformity in contemporary Iranian life \cite{sahebjame2012marriage}.

Existing research on romance scams highlights strategies like emotional grooming, staged intimacy, and anonymity to gain trust \cite{whitty2013scammers}. However, this work largely reflects Western contexts, where sex education and child protection frameworks offer structural safeguards. Little is known about how minors in restrictive environments—where sex remains taboo and early marriage is normalized—navigate online romantic advances \cite{cucci2019adolescent, hornor2020online, hornor2022online, muniz2019parental, Clarke_1980}. In Iran, these vulnerabilities are compounded by digital autonomy beyond parental oversight \cite{maloney2020virtual, johnson2010young, kumar2017no}, combined with cultural silence and the emotional weight of first love.
By foregrounding the lived realities of Iranian minors, we argue that online dating romance scams are not only technical or psychological issues—they are deeply entangled with sociocultural structures that shape youth agency, secrecy, and vulnerability.

\section{METHODOLOGY}
 This study employed episodic narrative interviews with 16 adults who experienced online dating romance scams while under the age of 18 in Iran. Participants were recruited primarily through an Instagram account themed around romance in Iran, and through personal networks Data were analyzed using qualitative content analysis combining inductive and deductive approaches.

\subsection{Participant Recruitment}
The primary recruitment strategy involved public posts on an Instagram account managed by the first author, which has 12,000 followers. This was not an account for the author's personal life, but rather one themed around romance in Iran by posting Persian-language content (e.g., love/break-up videos, texts, poems, and music). While this approach arguably biases the sample towards those already comfortable with discussing the study's subject matter openly, it is important to contextualize this choice in Iran where human subject research recruitment infrastructure is largely nonexistent, and research into sexuality in particular faces legal and political obstacles \cite{rahmani2015sexuality}. Relatedly, romance/dating information is taboo in Iran compared to Western counterparts; social media content like the aforementioned account represents one of the very limited ways through which young people in Iran can consume and partake in dialogue about dating and sex in Iran. Ultimately, the choice of a social media-oriented recruitment strategy was largely out of necessity, not convenience, due to limited alternatives.

Eleven of the 16 participants were recruited through posts about the study on the Instagram account; an additional five were recruited through personal connections to reach a point of data saturation in analysis while also being sensitive to inherent recruitment barriers and risk of participation for this study population \cite{marshall2013does, francis2010adequate, bekele2022sample, bowman2023using}. 
While this was a hard-to-reach population, our sample size of 16 nonetheless aligns with average sample sizes for interview studies published at CHI \cite{Caine}.
No monetary compensation was offered, which was a deliberate choice based on the authors' prior experience with human subject research with Iranian populations where participation was primarily driven by intrinsic motivation to contribute to research.

Our inclusion criteria required participants to self-identify as having experienced "online dating romance scams" as minors (under 18), through which they were deceived out of an object of value by someone with a fake identity. We received several inquiries from individuals who were simply heartbroken by an online relationship ending, but otherwise not scammed or deceived by someone with a fake identity; such individuals were not interviewed or included in the study.



Participants were aged 14 to 17 at the time of the scam, yet all were over the age of 18 at the time of interview. Participants described being scammed out of 1) money or material gifts, 2) sexual acts or sexual content, and 3) emotional investment; objects of loss that align with those in Whitty’s "persuasive model" \cite{whitty2013scammers} and Amirkhani's "body scam" model \cite{amirkhani2024beyond} of online dating romance scams. See Table \ref{tab:participant_overview} for an overview of participant demographics. See Table \ref{tab:victimization_types} for detailed reporting of the types of self-reported victimization through online dating romance scams. 




\subsection{Interviews }
We used episodic narrative interviews \cite{freeman1998experience, alizadeh2023catch}, an approach that emphasizes collecting individual stories within specific episodes to understand phenomena. As described by Mueller \cite{mueller2019episodic}, this method allows for a relatively open-ended exploration of experiences. The episodic framework has proven valuable in documenting the sequence of events surrounding crime, offering insights into offender strategies when perpetrating crimes, as well as the evolving victim experience \cite{cornish1994procedural, keatley2018crime}. 

Episodic narrative interviews for our study began by informing participants about the research objectives, confirming their consent, and prompting them to share their experiences with online romance scams using the open-ended question: "Tell me your story before, during, and after the incident." This format helped participants organize their narratives chronologically, providing a clear picture of the progression and impact of their experiences. Supplementary questions were posed afterward, if necessary, to clarify aspects that were not mentioned during the narration, such as: \textit{"Have you ever been in a relationship before?"} to check whether,  as a minor, it was their first romantic experience.

Interviews were carried out remotely using social media platforms (Instagram, Telegram, WhatsApp), allowing participants to choose between voice or text message formats based on their preferences for comfort and accessibility. This flexible approach was particularly beneficial given the sensitive nature of the topic, as it allowed participants to select the medium they felt was most private and secure. The duration of the sessions varied from 40 minutes to 2.5 hours, with longer interviews reflecting more extensive experiences (e.g., one participant whose relationship lasted four years). All interviews were conducted in Persian, the official language of Iran, to ensure clear communication and respect for cultural nuances. The interviews were transcribed verbatim and later translated into English by the lead author to maintain accuracy and consistency in the representation of the data.


\subsection{Ethical Considerations  }
Ethical approval for this study was obtained from the lead author's university ethics board. Beyond that approval, we 
implemented multiple safeguards to protect participants’ privacy, emotional well-being, and data security \cite{zheng2024s, chen2022trauma, amirkhani2023taking}. During recruitment, we relied on the first author’s public Instagram account and personal network because conventional research recruitment channels would have potentially been subjected to political and legal attention, especially given that premarital relationships are illegal in Iran.

All participants were assigned pseudonyms and identifying details such as real names and locations were omitted during interviews and subsequent analysis. 
Interviews were conducted online and adapted to participants’ preferences—via text or voice—often during late-night hours to ensure privacy and emotional comfort. Many participants chose these options to avoid being overheard by family members, to prevent potential stigma, or to keep their experiences hidden from current partners. To minimize the risk of re-traumatization, we adopted a trauma-informed approach grounded in established ethical guidelines \cite{campbell2019trauma, alexander2018systematic, chen2022trauma}. This included offering participants control over the pacing of the conversation, allowing them to pause or reschedule, and encouraging them to stop the interview at any point. We also provided participants who might feel traumatised by events with contact information for psychological support groups and professional services.  Data from each interview was stored securely on offline devices and participants had the option to delete chat histories. 

 After each session, the interviewer debriefed participants with questions such as \textit{“How are you feeling right now?”} and \textit{“Is there anything you need before we end the session?”} to identify if participants were in immediate distress. While none indicated a need forsupport at the end of their interview, three participants reached out after their interviews to inquire about professional psychological treatment related to their scam experience (not the interview experience); they were directed to a curated list of trusted mental-health professionals in Iran offering online counseling. To broaden access to preventive and coping resources, the first author facilitated two live-streamed interviews with a psychologist to discuss preventive strategies against romance scams as well as coping strategies for those who had already been victimized. Participants were informed about these sessions, which were also promoted on the Instagram account to all 12,000 followers and open to the public so that attendance in the live-streams would not imply participation in the study.

\subsection{Data Analysis}
We initially approached the data with qualitative content analysis (QCA) \cite{glaser2020potential, hsieh2005three}, supported by MAXQDA. However, the analytic process increasingly required interpretive engagement with cultural and relational meanings, particularly once collaborative coding commenced with multiple researchers who represented diverse Iranian and non-Iranian upbringings. As Iranian contextual dimensions became central to the analytic process, our approach shifted to reflexive thematic analysis (RTA) \cite{Braun2019}, which values researcher subjectivity and supports a combination of inductive and deductive coding, the latter allowing personal knowledge and literature about authoritarian parenting, adolescent online vulnerability, and other sociocultural issues in Iran to inform analysis.

RTA entails the following \cite{braun2021thematic}: 1) data familiarization; 2) initial coding; 3) generating initial themes by organizing codes; 4) developing and reviewing themes; 5) refining and naming themes; and 6) writing up results. Four authors actively contributed to the analysis. The lead author performed data familiarization (step 1) by creating transcripts of the interviews and translations to English, along with creating an initial set of codes (step 2) that were reviewed and discussed among the co-authors over multiple iterations (gradually shifting into step 3 of RTA with initial themes). When multiple coders are involved, RTA does not employ intercoder reliability, but rather collaborative coding \cite{braun2021thematic}, because it treats researcher subjectivity as a resource rather than a flaw to be identified and corrected. Collaborative coding was used reflexively: discussions among the researchers surfaced alternative interpretations, with Iranian authors offering cultural explanations essential for interpreting nuanced rural and urban sociocultural dynamics, and non-Iranian authors helping challenge assumptions and motivate elaboration on codes.
Because not all coders were from the Iranian context, the first author was often deferred to for explaining the \textit{“situated and contextual”} \cite{Braun03072021} dimensions of some participants' stories, which often necessitated elaborate explanations of Iranian culture and differences in rural and urban sociocultural dynamics.

Through recurrent team discussions, codes were iteratively organized into themes (steps 4-5) that captured individual experience of participants along with broader sociocultural dynamics. These thematic interpretations underscored how scammers leveraged cultural expectations around love, education, marriage, and autonomy in their grooming strategies. Consistent with both the exploratory QCA-inspired phase and our later shift to RTA, we did not develop a fixed codebook; instead, codes evolved iteratively without fixed definitions. The thematic structure was finalizsed through the writing of the Findings section for this paper (step 6 of RTA).

\section{Findings }
The overarching theme from data analysis was how online dating romance scams against Iranian minors in our sample exploit local cultural pressures imposed or reinforced by parents.
Our resulting thematic structure organized these cultural dynamics, and scammers' strategies for exploiting them, along a chronological axis, providing an anatomy, of online dating romance scams 
that participants consistently described (See Table~\ref{tab:scam_stages_image}).

\begin{table*}[t]
  \centering
  \caption{Anatomy of an Iranian online dating romance scam against minors.}
  \label{tab:scam_stages_image}
  \includegraphics[width=\textwidth]{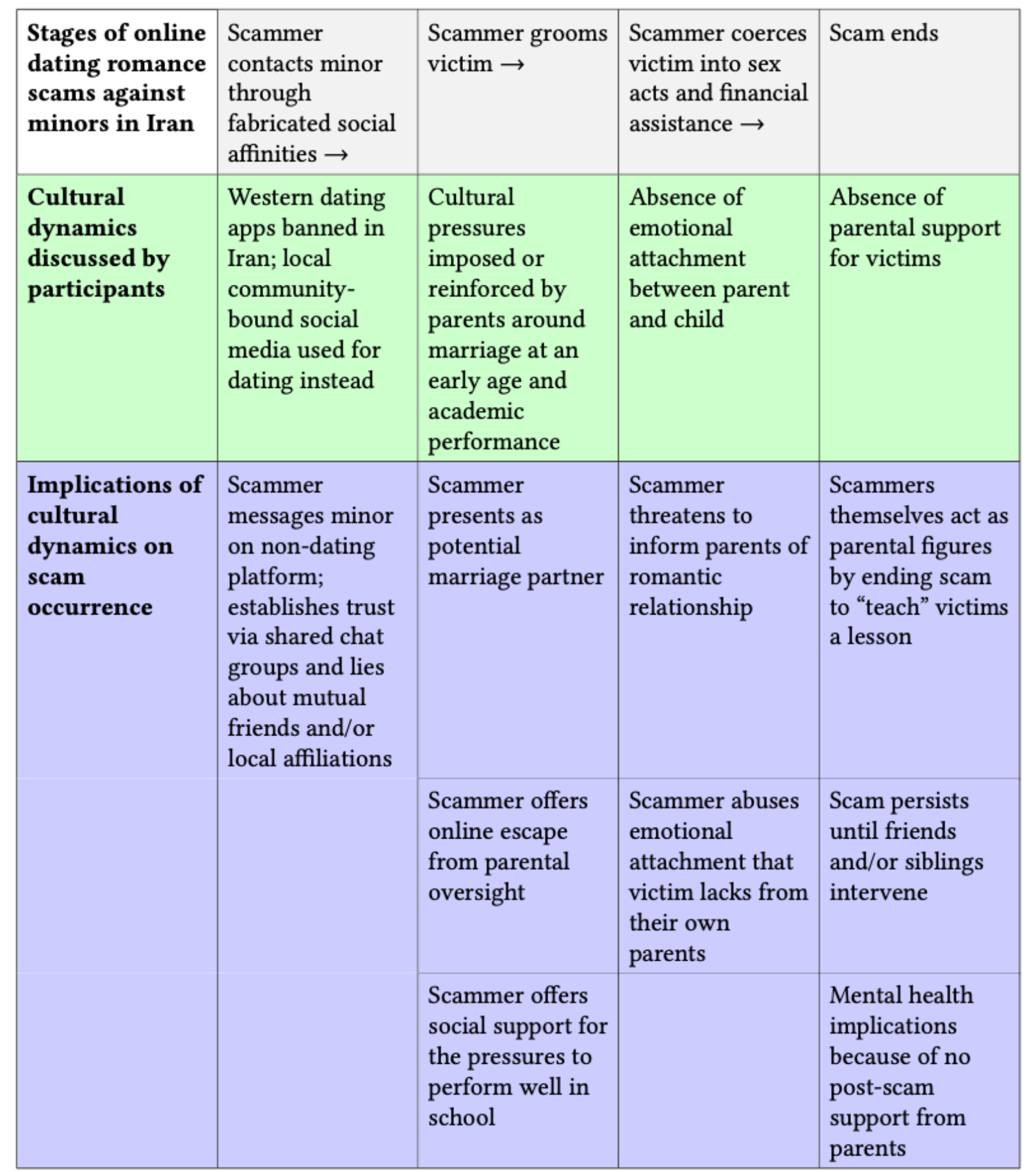}
\end{table*}

The process began with scammers sending private messages to minors on non-dating platforms commonly used in the local community such as Telegram because of Dating-app bans and everyday Instagram/Telegram use in Iran (\textbf{section \ref{4.1}}). Scammers then spent prolonged periods of time grooming their victims by offering relief from local cultural and parental pressures (\textbf{section \ref{4.2}}). These pressures included 1) expectations for marriage at an early age (and therefore a sense of competition that minors feel to procure a marriage partner quickly), 2) extreme parental oversight that instigates a desire for freedom and autonomy online, and 3) pressure to perform well in school. After building an emotional attachment, scammers manipulated victims into sending financial resources or sexual content by tapping into the same cultural and parental pressures (\textbf{section \ref{4.3}}). Such strategies involved threats to inform the minor's parents of their sexual relationship (which is taboo outside of marriage in Iran) and telling lies that minors easily fall for because of a (fake) emotional attachment that victims do not get from their own parents. The aftermath of the scams was also intertwined with cultural dynamics (\textbf{section \ref{4.4}}). Because participants lacked family support post-victimization they experienced poor school performance and frayed relationships with family members due to loss of trust.

In the following sections, we report in more detail on each stage of the online dating romance scam process and the intertwined Iranian cultural dynamics.

\subsection{Scammers Contact Minors on Non-Dating Platforms Through Fabricated Social Affinities} \label{4.1}

Romance scams in Western cultures typically begin with contact on dating apps, where users create profiles specifically for dating reasons, swipe on profiles of nearby daters, and exchanges messages under the explicit notion of dating. On the contrary, all of our participants in Iran described their scams originating on 
\textit{non}-dating platforms such as Instagram and Telegram. These platforms are deeply embedded in everyday social life in Iran, particularly for minors, and thus function as de facto dating spaces. Their use for dating reflects both practical and cultural conditions: dedicated "dating apps" popular in Western countries were either fully inaccessible in Iran or perceived as stigmatized, while Instagram and Telegram were widely accepted for maintaining friendships, community ties, and entertainment. 

Fake identities form the crux of romance scams. While such scams in Western cultures often involve the use of fake profile pictures (physical appearance), our participants most commonly described their scammers using fake social affinities such as mutual friends, shared regional affiliation, or participation in local group spaces. These fabrications had powerful influence on participants because the use of social media apps in Iran is closely intertwined with one's local geographic area (e.g., shared group chats for a given neighborhood are common). Scammers being able to tout affiliation with a participant's local community was thus a way to expedite trust and avoid being seen as a stranger.

Instagram provided a profile- and image-based environment where romantic interest could be expressed through follows, likes, and comments, often leading to private/direct message exchanges. Several participants described how casual interactions around shared images or social events were used by scammers to gain initial trust and develop the guise of a personal connection. 
For instance, Melisa described how her scammer built trust by referencing a supposed mutual connection related to a party which was later learned to be entirely fabricated: \textit{“I was invited to a party. After the party, some of us found each other on Instagram and created a chat group there. We usually talked about different cultures, traditions, and parties—it was just general conversation. Then someone messaged me privately, saying he was a friend of a friend who was also in the group.”} False shared connections were echoed by several other participants.

Amin similarly described how visibility on Instagram led to gradual attention from their scammer that shifted from public comments/likes on their content to private messaging: \textit{“My Instagram account was not private [it was visible to everyone]. I usually shared some photos of myself there. [The scammer] often replied to my photos on Instagram’s story and said she liked them.”} 

By contrast, Telegram centered on group chat participation, often tied to local communities, regions, and interests. Here, scammers could leverage a sense of local familiarity to initiate private conversations. One participant, Hiva, recalled how they met their scammer: \textit{“A chat group for our province. Both of us are in the same province but different cities.”} Similarly, Nina alluded to how the platform felt inherently trustworthy because their friends also used it: \textit{“It was a messenger application. I had some of my friends there.”} Multiple accounts illustrated how Telegram’s group-based dynamics created a context of perceived safety and legitimacy, from which private romantic conversations could emerge. For instance, Sheri explained how group-based connections could gradually shift toward private exchanges:

\textit{“I was on Telegram and suddenly I saw a link to a chat group and I joined. [...] I saw him in this group for the first time. He introduced himself and was from another city. We were friends like others, but after some months he started to text me in private chat, and at first it was still a normal and simple friendship.”} Sona also recounted a similar progression: \textit{“When I was in high school, on Telegram, by my friend, I was invited to a group. In this group I met a boy. He reacted to all of my conversations kindly. [...]. Then one of his friends texted me and told me he liked me.”}

These initial exchanges across non-dating platforms often made reference to courtship and marriage as a way to establish romantic intention. Alluding to marriage with a teenager may appear unusual or inherently predatory in Western countries, but such discussions were culturally normalized and significant for minors in the regions where our participants lived. Melisa explained: \textit{“I wasn’t looking to date. One day, someone messaged me on Instagram, complimenting my pictures and calling me pretty. Eventually, he said he wanted a relationship and even talked about marriage.”} Such references to marriage were not a cause for concern, but rather lent credibility and emotional significance to the interaction, aligning with local expectations about romantic commitments.

Over time, trust and intimacy were cultivated through extended, often late-night conversations, which gradually made the scammer a central figure in victims’ lives. As Nina reflected, \textit{“we chatted almost every night for around a year, often until 4 or 5 a.m.”} The emotional demands of these interactions were considerable, reshaping participants’ daily routines—\textit{“He filled all of my leisure hours.”} (Nina).

Through these interactions, scammers incrementally positioned themselves as socially proximate rather than anonymous—first as group members, then as acquaintances, and eventually as potential romantic partners. This gradual escalation allowed scammers to construct an identity that appeared embedded within the participant’s local social fabric, reducing suspicion and accelerating trust.

\subsection{Scammers Groom Minors by Offering Relief from Iranian Cultural Pressures Reinforced by Parents} \label{4.2}
Once scammers established initial connection and comfort with their minor victims—a stage that typically took several months—they entered a grooming stage through which they built (fabricated) romantic relationships with their victims online by tapping into sociocultural and geographically-specific vulnerabilities.
These vulnerabilities are not isolated factors, but are interwoven with broader cultural narratives around love, marriage, autonomy/authority, and education. While these pressures were shared across participants, the specific ways they were experienced—and exploited—
varied according to participants’ regional (rural/urban) contexts and socioeconomic backgrounds.  We identify three key, interrelated types of vulnerability that were targeted by scammers.

\subsubsection{Exploiting Expectations for Marriage at an Early Age}
 Our data show differing expectations between participants from rural and urban contexts. Participants from rural areas described marriage as a valuable achievement, reflecting how early marriage remains a marker of family honor and social stability in their neighborhood.  
For many families, a girl’s desirability is seen as a reflection of both her worth and her family’s reputation. Delays in marriage can invite gossip or suggest failure to uphold cultural norms. As Ziba described: \textit{“In our region, when a girl turns 13, the family starts pressuring her to marry. If no one proposes, it means she’s not lovable. [...] So when someone online told me he loved me and wanted to marry me, I was already in that mindset.”} Similarly, Hiva emphasized her goal of finding a marriage partner and how scammers exploited this desire: \textit{“I wanted to find an appropriate person for marriage […] he pretended he wanted me for marriage.”}

Ziba’s and Hiva’s accounts illustrate how scammers exploit cultural scripts around early marriage: romantic advances and promises of marriage become signs of validated self-worth rather than red flags. This dynamic reframes a scammer’s interest as social achievement rather than a threat, proving that the victim is desirable and worthy.

Participants from urban parts of Iran described a different, yet equally potent, form of social pressure; one driven not by family expectations for marriage, but by the peer economy of popularity. In these environments, having a boyfriend or girlfriend was not simply about personal choice; it was a public marker of social worth. Zahra captured this competitive logic when she described what being in a relationship felt like: \textit{“When all my classmates are in relationships [...] it was my turn to show that I am lovable.”} Her use of the word \textit{"show"} was indicative of a recurring theme about relationships operating as a currency of status among peers, where romantic attention validated not only attractiveness but also belonging within the group.

For boys, the same dynamic applied, though mediated by norms of masculinity and confidence. Tiam recalled feeling inadequate compared to classmates who were already in relationships: \textit{“All my classmates already had girlfriends. But I was too shy to ask a girl out. I always hoped someone would come to me.”} His hope that a relationship would arrive without his initiating it hints at a vulnerability scammers could easily exploit—offering what seemed like a stroke of luck rather than manipulation. When such attention appeared in a Telegram group, Tiam interpreted it as an overdue opportunity to restore his social standing.

These accounts suggest that scammers did not operate randomly; they embedded themselves within existing systems of value. By framing their advances as evidence of desirability or popularity, they tapped into the emotional economy of teenage peer culture in Iran where being chosen signified validation. These accounts suggest that, unlike overt deception, these strategies mirrored legitimate social scripts, making them harder to detect and resist.

\subsubsection{Exploiting Parental Oversight and Minors' Desire for Digital Freedom}
Strict parental oversight was a defining feature in many participants’ lives; rules that were meant to protect often translated into rigid control. 
Lili recalled being \textit{“under immense parental pressure,”} describing how these restrictions left her with little room to socialize or even meet friends: \textit{“At that juncture, I was under immense parental pressure. My parents’ strict oversight restricted my social interactions and prevented me from freely engaging with friends or going out.” (Lili)}
Such boundaries, while framed as safeguards, created a paradox: the more tightly parents monitored the physical world, the more appealing the digital one became.


Sarah articulated this tension vividly, noting that\textit{“although everyone here is closely controlled in the real world, they haven’t yet managed to control the virtual world.”} For her, cyberspace was both a loophole and a lifeline—a space where surveillance faltered, offering freedom that felt empowering but also exposed her to risk. This sense of liberation was echoed by Hiva, who admitted she went online \textit{“just to have fun,”} underscoring how ordinary motivations, rather than calculated defiance, drove many toward digital spaces.
Others, like Sheri, described how scammers reinforced this feeling of empowerment by offering emotional guidance:\textit{ “he helped me mentally and emotionally to [...] be brave enough to talk with my parents.”} she recalled. What seemed like mentorship was, in reality, manipulation through the appropriation of the language of care and independence to strengthen trust.

Scammers recognized this vulnerability and exploited it with precision. They framed themselves not as intruders, but as allies granting access to the autonomy minors craved. For Lili, discovering a way to connect with someone on Telegram without her parents’ knowledge felt like reclaiming agency: the thrill of doing something unsupervised blurred the warning signs: \textit{“When I discovered the opportunity to connect with a boy on Telegram without parental oversight, I readily embraced it.”} 
Moreover, she described how even small gestures of concern helped \textit{“relieve the mental pressure,”} signaling that scammers’ strategies were effective because they filled an emotional gap parents often overlooked. In these stories, care was not romanticized fantasy but a practical antidote to chronic stress. Per Lili, \textit{“It helped me feel calm and relieved the mental pressure [...] due to family issues.”} 


Ultimately, these stories reveal a striking contradiction between control and secrecy. The very mechanisms designed to shield minors from harm pushed them into online spaces hidden from their parents where scammers could present themselves as liberators, validating identities and desires that families left unacknowledged.

\subsubsection{Exploiting Academic Pressure and Need for Emotional Support}

In contrast to an early marriage emphasis reported by participants in rural areas, participants from urban areas and more liberal families in our sample described a familial prioritization of academic achievement as a path to social mobility and future job security. Education, especially college admission for competitive fields like medicine, 
 was deeply tied to family reputation and long-term job security. Education in these families was not only a personal goal but a collective moral duty, framed as both a family investment and a safeguard against economic instability for all family members beyond the immediate child that receives the education \cite{zonooz2021employment}. This pressure, while intended to guarantee a \textit{“safe future,”} became a vector for grooming by scammers.  

Participants like Sheri grew up in families who prioritized educational achievement as a point of family pride and were deeply involved in their children's lives. 
Instead of buffering stress, parental involvement often amplified it, leaving little space for emotional support. One example is pushing children toward highly competitive academic programs through the Konkoor (the university entrance exam in Iran), especially in fields like medicine. Sheri shared her experience:\textit{“I hate the medical field. I had to put myself under a lot of pressure to achieve something I never wanted. My parents never considered my interests. They told me that in Iran the only way to have a secure future is to work as a medical doctor. This made me feel stressed, like I was dying.”}

Sheri’s account illustrates how academic success was 
 transformed from opportunity into obligation, exemplifying how educational pressure in Iran’s moralized, family-centered context differs from individualistic notions of academic success found in other (particularly Western) contexts.

These intense academic expectations, compounded by emotional neglect, created fertile ground for scammers to intervene. Participants’ accounts revealed how scammers offered emotional support as a form of refuge from the stress of family-imposed academic goals. Sheri elaborated on how a proposed romantic relationship with her scammer was a much needed relief: \textit{"My parents’ [academic] expectations were overwhelming, adding immense pressure. At that time, I felt depressed and stressed. Accepting his love-friendship suggestion eased my stress because of the attention and care he gave me.”} This quote illustrates a critical theme in our analysis in that academic pressure was not only exhausting, but isolating, making emotional validation feel like rescue rather than risk.
 
For some, scammers’ tactics extended beyond emotional comfort to instrumental promises. Nina recalled: \textit{“I was studying hard to get ready for the Konkoor. He suggested starting a romantic relationship, promising to help me study and offering emotional support. I thought it could be a good chance for me to gain love and help from someone who cared for me.}” Nina's scammer appeared to understand how crucial the Konkoor was for adolescents’ futures; by promising help with such a high-stakes exam, they made their affection seem both emotionally meaningful and practically invaluable. For participants who viewed "Konkoor" success as a rare pathway to socioeconomic stability, offers like these transformed the scammer into someone seemingly indispensable, reinforcing dependency at a moment of acute vulnerability for high schools students nearing graduation.

Parental expectations for academic excellence also amplified the importance of leisure and recreation to minors. Short periods of academic respite, such as post-exam breaks or summer holidays, were exploited by scammers as an opportunity to build further emotional investment while minors were spending more time online. 
 
Sheri noted turning to social media \textit{“with plenty of free time after exams,”} while Arman admitted that during holidays, \textit{“it was summer, no school, no hobbies.” } 
Taken together, participants' narratives reveal a recurring pattern: relentless academic demands stripped minors of autonomy and emotional support, while unstructured leisure heightened their online activity. Scammers capitalized on both conditions by providing emotional and even practical support, positioning themselves as uniquely supportive at moments of heightened vulnerability.

\subsection{Scammers Coerce Minors into Sexual Acts and Financial Assistance Through Parental and Cultural Pressures} \label{4.3}

 Participants described how scammers gradually incorporated dynamics of manipulation, fear, and obligation into their relationship, manifesting in exploitative demands that typically involved repeated sexual and/or financial favors.
 
    \textit{Sexual Coercion:} Several participants reported sustained pressure to provide sexual content online. Mahi’s account illustrates this trajectory in which initial requests for photos without a hijab escalated into demands for fully nude images:\textit{“He requested a photo from me without my hijab. […] After receiving some of my photos, he began pressuring me for naked photos.”}  
 These narratives reveal a consistent mechanism in which scammers first establish emotional rapport, then exploit intimacy to erode participants’ privacy and bodily autonomy, creating a sense of inevitability and helplessness.

   \textit{Financial Exploitation:} A second pattern involved manipulation for material gain.  
   Melisa described an experience, though without sexual coercion, of providing small amounts of money for online purchases or phone recharges to a female scammer impersonating a man: \textit{“He, which was actually she [a female scammer pretending to be a man], never asked me for sexual acts. [...] I sent her small amounts for online purchases or phone recharges.”}  
   
\textit{Combined Sexual and Financial Pressure:} In some instances, scammers intertwined sexual and financial demands. For example, Maria recounted performing sexual acts on video calls while simultaneously providing monetary support over two years: \textit{“He made me perform [sexual acts censored by authors] on video calls, and I complied out of fear and shame. Ridiculously, I even had to recharge his phone for internet access, sending him small amounts of money over two years. I felt trapped and disgusted.”} This dual exploitation underscores the multilayered control exerted by scammers, where compliance in one domain reinforces vulnerability in the other. 

Scammers' tactics for receiving sexual and financial favors exploited the same parental and cultural dynamics previously mentioned, with strategies involving fear of parental judgment, lies that capitalized on minors' dating inexperience, and manipulations of emotional attachment.

\subsubsection{Threats to Reveal Taboo Sexual Aspects of Relationship to the Minor's Parents}

Some scammers maintained control over minors by leveraging the threat of parental exposure to sexual aspects of the relationship (unlike marriage itself, engaging in sex acts outside of marriage is considered taboo).
In contexts where parents are perceived primarily as enforcers of honor rather than sources of emotional support—especially in rural areas—disclosure of sexual acts carried the risk of blame, shame, or punishment. This created a coercive environment in which the possibility of parental discovery functioned as a powerful instrument of compliance.

Mahi’s account, for instance, illustrates how the threat of sending private images to her father transformed resistance into compliance: \textit{“He threatened to send my pictures to my father if I didn’t comply with his demands.”} Lina echoed this dynamic, emphasizing that fear of parental judgment prevented her from seeking help: 
\textit{“I couldn’t discuss it with my family because I was certain there would be no support for me, and they would just blame me.”} Hiva further described the emotional calculus of silence, weighing the anticipated barrage of shaming messages against the uncertain safety of parental intervention, and ultimately choosing secrecy as the lesser risk:\textit{“Instead of receiving numerous shaming and blaming messages, I preferred to keep it a secret because I did not feel safe about my father’s reaction.”}



These threats consistently produced two intertwined outcomes: sustained victim silence about the now-revealed scam and continued compliance with the scammer's demands, whether through sharing sexual content or providing financial support. The pattern underscores a broader mechanism of coercion: scammers exploit pre-existing vulnerabilities rooted in cultural expectations of honor and authority, transforming minors’ fear of social and familial repercussion into a tool for manipulation.

\subsubsection{Exploiting Emotional Attachment with Lies and Threats to End the Relationship}

Scammers systematically leveraged minors’ emotional vulnerabilities, including their desire for affection, attention, and stress relief, to establish a false sense of intimacy and dependence. Emotional reassurance functioned as a coping mechanism that participants sincerely believed was reducing psychological distress. For example, Lili reflected that \textit{“his support and attention with love decreased the mental pressure and made me more calm,”} highlighting how even minor gestures of care were sufficient to anchor her emotionally to the scammer. Amin provided a more elaborate explanation of how emotional attachment was leveraged whenever they became suspicious that the relationship might be a scam. 
\textit{“Whenever I was curious, she threatened me if I stayed suspicious, she would cut the relationship. Since I loved her, I stopped being curious. My dependence did not let me behave logically. It was my first love, and I was so dependent on her attention and love bombing.” (Amin)} 

The combination of affection and subtle threats exemplified in Amin's story shows how effectively the strategy constrained victims’ autonomy. Amin described being warned that \textit{“if I stayed suspicious, she would cut the relationship,”} a threat that, coupled with first-love attachment, made him suppress doubts and ignore red flags.

Similarly, Sheri’s account shows how empathy and love dependency prolonged involvement even when suspicion is valided by discovering proof of the scammer's dishonesty.
\textit{“When I discovered through a blog that he was using a fake photo… his claims about being ‘ugly’ and using the photo out of insecurity kept me hooked.” } Here, manipulation of empathy and perceived vulnerability prolonged the emotional entanglement. 

Scammers also capitalized on minors’ limited dating experience and naivety, 
constructing narratives that appeared plausible and non-threatening. Simple lies, such as Sheri believing \textit{“his phone camera was broken,”} prolonged scams for years in some cases because victims lacked the relational or technological experience to detect inconsistencies. 
Participants acknowledged their own inexperience as a contributing factor. For instance, Nina reflected, \textit{“I didn’t do it intentionally, [...] I simply lacked experience,”} while Lili noted, \textit{“When I was 17, I had no idea how much of a bad experience it could be to date the wrong person in cyberspace.”} 
By intertwining affection, emotional manipulation, and the threat of relational loss, scammers created a self-reinforcing cycle of dependency. 



\subsection{Implications of Lack of Parental Support on the Ending and Aftermath of Scams} \label{4.4}


No participant felt safe disclosing to their parents that they were in a romantic relationship with their scammer, even when they believed the relationship to be legitimate and progressing towards marriage. This was due in part to engaging in cybsersexual activities (taboo outside of marriage) and parents generally being perceived as a source of pressure and expectation rather than support.

Romance scams thus could persist for months or even years without a parental authority to identify signs of concern. For several participants, it was the scammers themselves who ended the relationship, 
framing the willful disclosure of the scam as a lesson or moral guidance about online safety. Maria described how her scammer developed genuine feelings for her, which drove the scammer to admit to the scam and end the relationship: 
\textit{“One day, he claimed he liked me and said his actions were to teach me not to trust strangers. He told me to break my SIM card to end the relationship, and I did so immediately, finally escaping the nightmare.”} For Tiam, while the scammer did not directly acknowledge the scam, their sudden disappearance after receiving money all but confirmed that the relationship was fake: 
\textit{“During the time we were in a relationship, I never suspected. When she disappeared after the money scam, I understood I was a victim. I could not do anything about it, nor did I want to.”}



This subsection examines the implications of absent parental support on the ending and aftermath of romance scams. First, we look at the role of siblings and close friends as informal guardians who perform vital interventions. We then elucidate the mental health impacts on victims in the aftermath of scams that are triggered or exacerbated by the absence of parental support.


\subsubsection{Scams are Prolonged Until Intervention by Siblings and Friends} 


Support from siblings and friends played a crucial role in helping participants recognize and terminate the scam. Whereas parents were seen as an unsafe option for support, several participants confided in siblings and close friends about suspicions of a scam and their supposed-romantic relationship more generally, whose advice was often the catalyst to recognizing the relationship as a scam and  ending it.

For example, Mahi described how her older sister identified the coercion and confronted the scammer on her behalf:  \textit{“I sought assistance from my sister, who, being a decade older, recognized the pressure I was facing. She confronted him [the scammer] and warned that any further misconduct would lead to legal action.”} Siblings were in a unique position to provide support because they lived in the same household and thus could more readily spot concerning relationship patterns, albeit without being associated with the pressures and expectations of parents that might render a scam victim hesitant to confide in them.

Close friends complemented this support by offering emotional guidance and practical strategies for disengagement. Mehri described the combined support of her sister and close friends in helping her end the scam: \textit{“My sister gave me hints to be more aware of what I was doing, and when I went to university, my friends supported me mentally, helping me cut off the relationship and cope with losing a fake love.”}  Similarly, Melisa’s friend encouraged reporting the scam to authorities and helped her manage feelings of shame:
\textit{“When she [the scammer] avoided video calls and disappeared for 19 days, I realized it was a scam. When she returned asking for money again, my friend urged me to report it to the police, reassuring me not to feel ashamed since no sexual act had occurred.”} 




\subsubsection{Mental Health Impacts Due to Absence of Post-Scam Support from Parents}

The absence of robust parental support in the aftermath of romance scams had significant and lasting consequences. Participants described enduring psychological distress, including anxiety and insomnia. 
Sona described declines in academic performance and sleep: \textit{“There were a lot of bad impacts on me, ranging from poor academic performance to insomnia.” } In the most severe case, the psychological toll led to multiple suicide attempts. Hiva recounted: \textit{“I made ten attempts at suicide over the course of a year, but none of them were successful.”} When asked why she felt driven to such extreme actions, she noted that her fears were shaped by local cultural expectations surrounding premarital romantic interactions, and how anticipated cultural consequences of her scam seemed worse than harm she could inflict on herself. In her words: 

\textit{"I could imagine that in our conservative region, after having sex, I would have no chance for marriage. I could not feel safe around the male members of my family—my dad and brothers—so I preferred to die in an honorable way rather than live a life without prospects and face harsh punishment from male family members."}

These experiences illustrate how mental health impacts in the wake of romance scams are predicated not only on the scams themselves, but \textit{"conservative"} cultural perceptions of premarital romantic or sexual activity. Participants anticipated blame or punishment rather than support, which made disclosure feel unsafe and intensified psychological distress.
These cultural norms carry long-term social repercussions, including diminished marriage prospects, as reflected in Hiva’s fear of facing a \textit{“life without prospects.”} At the same time, the collectivist structure of Iranian families meant that perceived transgressions threatened not only the individual’s reputation but the family’s. Participants described both physical consequences (e.g., Hiva’s fear of \textit{“harsh punishment from male family members”}) and emotional ones, such as deteriorating trust and support. As Lili noted,
\textit{“After the incident, my mother no longer trusted me, which caused additional mental stress and a decline in my studies.}”

The notion of dishonoring the family led to disapproval not only of one's parents, but siblings as well in some cases (demonstrating how siblings played drastically different roles in emotional support or duress across our sample). Zahra explained 

that her sister’s anger and emphasis on high expectations made her feel morally and socially inadequate: \textit{“She was so angry when she found out I was in a relationship with such an ambiguous guy. She told me she had much higher expectations, and that if our parents found out, they would completely lose their trust in me.”} 
Tiam also spoke to worsened relationships with the broader family: 
\textit{“The important thing is all of my family members look at me like a stupid person. After this experience they never trust me anymore. And I am sure I cannot catch their trust again.”}

These accounts collectively demonstrate that online romance scams do not merely produce temporary emotional or financial harm; they can generate long-term psychological distress intertwined with familial tension. 
In cultural contexts where family reputation and obedience are highly salient, the perception of having failed one’s family can intensify trauma, creating a dual burden: the direct effects of exploitation and the indirect consequences of perceived social and familial judgment.

\section{DISCUSSION}
In this paper, we extend online safety research by examining minors as \textit{victims of online dating romance scams 
in Iran}, an intersection rarely studied. Prior work on minors' online risks has largely focused on cyberbullying, exposure to explicit content, or sexual solicitation and grooming \cite{hamm2015prevalence, giumetti2022cyberbullying, chen2019doxing, alsoubai2022friends, lee2022phishing}. Research on romance scams, in contrast, has centered on adults—mainly in Western contexts \cite{ortega2022epidemiology, hsieh2023understanding, hornor2022online, hsieh2024correction, giumetti2022cyberbullying,hamm2015prevalence, kopczewski2018cyber}. While studies on adolescent dating violence exist \cite{muniz2023victims, muniz2017online}, they emphasize offline, physical relationships or frame minors as perpetrators \cite{hinduja2008cyberbullying}, overlooking the technology-mediated, relational, and financial harms when minors are victims of scams.  
 To do so, we first articulate how online dating romance scams targeting minors relate to, overlap with, and diverge from grooming and sexual solicitation. We then examine how these scams are intertwined with Iranian cultural dynamics and parental expectations before turning to the downstream impacts and implications for digital resilience.

\subsection{The Relationship Between Online Dating Romance Scams and Technology-Mediated Harms Targeting Minors} 
 The defining quality of online dating romance scams is the use of a fake identity to cultivate a false romantic relationship online \cite{whitty2015anatomy, buchanan2014online, whitty2012better}.  The goal of romance scams is to extract "items of value" from the victim \cite{garrett2014exploring}, which are often financial in nature \cite{whitty2015anatomy}, but are also known to involve sextortion to solicit sexual content from the victim \cite{cross2023pay, cross2024if}.

 Three features of online dating romance scams in our findings depart from prior literature and expand the conceptual boundaries of romance scams. First, while the romance scams reported in our study did manifest through online dating, scammers and victims did not meet through traditional, Western-oriented dating apps, which are banned in Iran. Rather, scams developed through more general social media platforms like Telegram. This distinction is key in light of the second point of novelty: victims in our study were minors, not adults, who would not pass the minimum age requirement for Westernized dating apps even if they were not banned in Iran and thus could not be a population study in Western-centric romance scam studies. Prior work has overwhelmingly profiled victims who are middle-aged adults, and from \textit{"Western societies"} such as the United States (e.g., \cite{coluccia2020online}). The representation of middle-aged victims is likely due to a focus on financial fraud-oriented romance scams in the literature \cite{whitty2015anatomy, whitty2013scammers, suarez2019automatically, whitty2012online}, particularly dynamics in which Africa-based scammers deliberately target relatively affluent victims in Western countries \cite{eseadi2021hello, edwards2018geography}. The emphasis on sexual content exhibited in the romance scams from our study does echo prior discoveries about sextortion in romance scams \cite{cross2023pay, cross2024if}, and thus renders it unnecessary to coin a new term for sex-motivated romance scams. However our findings are notably distinct in the prioritization of sexual content as the primary or even exclusive focus on some romance scammers.

The confluence of  victims who are minor and exploitation for sexual content reported in our study emphasizes latent interconnections between romance scams and harms against minors, particularly child grooming \cite{craven2006sexual} and sexual solicitation of children \cite{hornor2022online}. These phenomena are not mutually exclusive with romance scams against minors, but rather essential stages \textit{within} the romance scam process. Grooming has consistently been depicted as a stage within online dating romance scams \cite{dove2024grooming, whitty2013scammers} to develop trust and (false) emotional connection with victims to, ultimately, increase compliance with eventual requests for the desired items of value from the victim. The findings in our study similarly position grooming as the second phase in the romance scam process against minors in Iran. Likewise, the requests (and direct demands) for sexual favors and material from victims later the romance scam process, as reported in our findings, align with definitions of sexual solicitation of children. Per Hornor and colleagues: \textit{"Online sexual solicitation occurs when children or adolescents are asked to engage in sexual activities, sexual talk, or give sexual information on the Internet"} \cite{hornor2022online} (see further support in \cite{mitchell2007youth}). This clarification of the roles of grooming and sexual solicitation within romance scams lays a foundation for exploring how culturally-rooted vulnerabilities play a role in each of these stages, which we reflect on in the next subsection.

\subsection{The Relationship Between Online Dating Romance Scams and Cultural Vulnerabilities}

Across participants in our study, secrecy and fear of parental judgment emerged as core features of their experiences. These shared patterns, which we used to confirm data saturation, show how deeply cultural norms around sexuality, parental authority, and online behavior structure minors’ romantic lives and scam experiences.
Some research like that of Alizadeh et al. \cite{alizadeh2023catch} has started to theorize the connection between romance scam victims’ vulnerabilities and offenders’ strategies in online fraud: \textit{"The attacker seeks to discover vulnerabilities in the “system” in order to exploit them selectively. In the case of SE [social engineering], these are typically human weaknesses that are exploited."} (P32:15) Moreover, as mentioned by Button and Cross \cite{button2017cyber} \textit{“It is up to the offender to decide on how to best match a potential victim with an appropriate fraudulent pitch.”} Our study shows how culturally-instilled vulnerabilities do not simply render minors susceptible to romance scams in Iran; they actively shape scammers’ strategies.

Cultural pressures surrounding marriage and romantic validation were consistently exploited through, for example, promise of marriage.
This intertwining of romance scams with cultural vulnerabilities resonates with research on sexual and romantic scripts, which highlight how culturally learned expectations around pursuit, desirability, and commitment can normalize persistence in sexual advances and obscure coercion into sexual acts \cite{greer2025persistence,simon2003sexual}. There are three specific cultural vulnerabilities showcased in our study that connect with prior work.

\textbf{Family-based fears:} Although studies show that parenting styles can significantly influence online risk by shaping adolescents’ psychological development and online behaviors \cite{simons2007linking, buelga2015family, muniz2019parental}, to the best of our knowledge the literature has not addressed scammers’ use of threats of exposure particularly to parents. Prior work does report on threats of sharing intimate images with peers or romantic partners as a coercive strategy though \cite{finkelhor2022prevalence,vogels2022teens, patchin2023digital, kowalski2014bullying}.
  
In the Iranian context, this suggests that parents are perceived less as sources of support and more as sources of risk. This tactic intersected with cultural norms in which disclosure could trigger shame, punishment, or loss of family honor—making silence appear to be the safest option.

\textbf{Naivety and optimism:} As noted in previous research \cite{muniz2017online}, minors have developmental vulnerabilities that exacerbate their susceptibility to manipulation and exploitation \cite{huh2023help, maloney2020virtual, stoilova2021investigating}. Our findings show that scammers exploited this naïve vulnerability by relying on fabricated stories and technical excuses that victims—due to their lack of experience—rarely questioned. Victims often internalized these lies as plausible.

\textbf{ Attachment needs:} As other studies show, adolescents whose leisure time is unsupervised or neglected are at greater risk of problematic Internet use \cite{lin2009effects}.
Our findings also highlight that unmet leisure needs increase minors’ risk of online victimization. Scammers capitalized on unmet needs for attention, affection, and relief from stress, fostering dependency through love-bombing and conditional reassurance. Even when suspicions arose, emotional entanglement kept victims compliant.

By exposing this intertwined dynamic, our study advances theory on online victimization in two ways: (1) by positioning minors as uniquely situated at the intersection of cultural scripts and digital affordances, and (2) by showing how scammers’ strategies adaptively map onto these vulnerabilities, forming a feedback loop of harm. Addressing this entanglement demands interventions that combine technical safeguards with cultural and developmental insights.

\subsection{ Potential Solutions to Romance Scams Against Minors}

Our findings portray a clear need for solutions to better prevent and manage the aftermath of romance scams against minors. Prior work about online harms against minors, while not necessarily situated in the Iranian context or romance scams in particular, could potentially inform solutions to romance scams against minors.

One approach is AI-driven content moderation \cite{pour2023comprehensive,borj2023online, wolbers2025artificial, ali2025artificial}, which could identify patterns of grooming behavior or other precursors to child exploitation. However, there are known issues with dataset quality in related forms of AI for sexual risk detection \cite{mandloi2024ai, islam2025artificial} that limit  effectiveness even in Western cultures where recognition of harms against youth is more well known than in Iran. These limitations are likely to be exacerbated in the Iranian context given non-English dialogue that render datasets in native languages rare or nonexistent. The popularity of platforms in Iran that are well known for the absence of moderation, such as Telegram, also raise concerns about whether a hypothetically reliable scam detection AI would even be implemented.

Another approach could be education for youth about online safety \cite{jones2014systematic,qadir2024towards}, which may introduce minors to romance scam risks and related harms in formal settings along with personal strategies for mitigation. Yet education-oriented solutions may incur sociocultural barriers to implementation \cite{oguine2025online}, not least because of the minimal empirical knowledge about such risks specifically in Iran.

A third approach could take the form of legal and regulatory measures \cite{USA_AdamWalsh_2006, USA_PROTECT_2003, Australia_OnlineSafety_2021, UK_SexualOffences_2003, EU_Directive_2011}. Examples in Western contexts include the U.S. PROTECT Act, which criminalizes online enticement and grooming of minors, and  the U.K. Sexual Offences Act 2003, which prohibits sexual communication with children. 

These legal standards, or similar frameworks, are likely to be very slow to implement in Iran and some other non-Western countries, in light of conservative laws regarding adjacent harms such as rape \cite{Iran_PenalCode}. For instance, Iran’s Penal Code requires four adult male witnesses to substantiate rape allegations, and Articles 102 and 638 criminalize “illicit relations,” which may result in punishment of victims. 



While our findings revealed a variety of vulnerabilities to, and devastating outcomes of, romance scams against minors, there were also early signals of resilience from our participants that could serve as a basis for alternative solutions. 
We use the term \textit{digital resilience} to describe the capacity to anticipate, withstand, and recover from online harms, while maintaining psychological and social well-being (Livingstone et al. \cite{livingstone2009kids}). Support from older siblings and peers emerged as a critical protective factor, functioning as both early-warning systems and emotional scaffolds. Peers provided reassurance and practical advice, including encouraging police reporting and reframing shame. These informal interventions illustrate how proximal social ties can serve as ad hoc safety nets in the absence of institutional mechanisms. Additionally, some minors attempted self-protection by breaking SIM cards, disengaging from platforms, or seeking distance from triggering environments. While these strategies indicate resilience, they also reveal the heavy burden of responsibility placed on minors themselves, often without structural or platform-level support, namely the scarcity of support services for \cite{cross2019you}.

A path to developing novel and effective resilience-oriented solutions could be through participatory approaches with youth that foreground their personal experiences in ideation of safety structures. This has been exemplified in HCI literature with some success \cite{ali2025teens}. While similar methods could be explored in the Iranian context, the risks should be considered of direct participation of minors from regions where research and the topic of study may be stigmatized. Per trauma-informed computing principles \cite{chen2022trauma}, for example, secondary stakeholders with indirect familiarity of the subject matter \cite{zheng2024s} in case this, romance scams and/or the lives of minors) could be approached for involvement in lieu of minors themselves.

\subsection{Limitations and Future Research}

\textbf{Memory recall:} One limitation of our study was the choice, for ethical reasons, not to interview minors directly after their scam experience (while they were still under the age of 18) when recall would be highest. We instead interviewed adults reflecting on their experiences from their adolescent years. While this decision was made for ethical reasons, we acknowledge how this limits recall fidelity and may have introduced retrospective rationalization or social reframing of events. Participants’ narratives, and findings from our study, should be interpreted as subjective reconstructions of lived experiences rather than objective timelines. 

\textbf{Participant verification: }Furthermore, we acknowledge that participants' alignment with our inclusion criteria, and particularly age at time of their romance scam, could not be fully verified, for practical reasons but also because  we did not want to impose a burden on victims of “proving” their sexual victimization in light of historical barriers to under-reporting of, and justice for, sexual abuse related to disbelieving victims \cite{ullman2023talking}. However we find the risk of falsified eligibility criteria to be reduced in our study because there was no financial incentive for participation. 

\textbf{Sampling:} In addition, as noted in the Method (Section 3.3), recruitment was conducted through the first author’s public Instagram account and personal networks. This strategy was largely a necessity for reaching this hard-to-access population due to challenges with formal research recruitment infrastructure and political and legal pressures, yet it also may over-represent individuals active on social media and comfortable acknowledging scam experiences. Future work may consider involvement of secondary stakeholders or less invasive methods such as surveys to expand representation of scam victims.

\textbf{Transferability: }Our findings seek not to be generalizable, but transferable because qualitative research recognizes that context is essential to the interpretation and meaning of resultant insights. Per Braun and Clarke \cite{braun2021thematic}: \textit{"Transferability refers to qualitative research that is richly contextualised in a way that allows the reader to make a judgement about whether, and to what extent, they can safely transfer the analysis to their own context or setting."} 
Because grooming practices, parental authority, and adolescent autonomy are culturally situated, future work should more directly examine the transferability 
of our findings to broader samples of Iranians, particularly to explore the differences between relative prioritization of academic success and early marriage exhibited between our rural and urban participants (which may not be transferable to all parts of Iran). 

Future work could similarly study other non-Western contexts beyond Iran, such as those with similar collectivist academic pressures. Although our interviews were conducted with Iranian minors, the mechanism we identify—exploitation of honor-related cultural scripts combined with adolescents’ developmental vulnerabilities—aligns with findings from other collectivist contexts where stigma inhibits disclosure (e.g., \cite{salhi2013gender}). Comparable mechanisms may manifest differently across cultures: in conservative societies through honor and marriage scripts, and in more liberal contexts through peer-driven competition or body image norms amplified by social media \cite{alluhidan2025unfiltered}. Moving forward, it would be beneficial to conduct longitudinal studies to further explore the long-term effects of society, family, and school support on minors' resilience post-victimization in online dating romance scams.

\section{CONCLUSION}

Our findings contribute to the online dating romance scams literature through a minor-focused lens by 1st person interviews, thus gathering nuanced, rich, and sensitive material which would otherwise be difficult to obtain. Also, we believe this study is innovative in offering insights into online dating romance scams targeting minors and their connection to societal and cultural factors—an area that remains largely underexplored.

Our study, we suggest, has unique elements. It advances the field by identifying previously under-explored factors that contribute to minors’ vulnerability in online romance scams by conceptualising experience as taking place in four stages: 1) Scammers messages minor on non-dating platforms with relevance to the local community
 2) Scammers groom minors by offering relief from Iranian cultural pressures reinforced by parents. 3) Scammers coerce minors into sexual acts and financial assistance through
parental and cultural pressures, 4) Implications of lack of parental support on the ending and aftermath of scams. By moving beyond reactive moderation toward holistic resilience infrastructures, we can shift from simply mitigating harm to empowering minors to navigate digital intimacy safely and confidently—while acknowledging the cultural and relational realities that shape their vulnerability.

\begin{acks}
This work was supported by the German Federal Ministry of Education and Research (BMBF) as part of the AntiScam project (Grant No. 16KIS2214). It was also partially supported by the U.S. National Science Foundation under Grant Nos. 2211896, 2401775, and 2339431.

Due to this research being situated in a non-Western context, English is not the native language for some authors of this work. We acknowledge the use of ChatGPT for copyediting portions of this paper's text for grammatical correctness, word choice, and for clarity of concepts when co-writing formative drafts of this paper across the author team, some of whom do and do not speak English as a first language. 

We thank Zivar Sheibani, a social teacher in Iran, for her valuable consultation during this work.

\end{acks}

\bibliographystyle{ACM-Reference-Format}
\bibliography{References}

\appendix
\section{Appendix: Narrative Interview Guide (Translated to English)}
\subsection{Step 1 – Narrative Prompt}

Main question (free narration):
\begin{quote}
    “Please tell me your story before, during, and after the incident.”
\end{quote}
\paragraph{Participants are encouraged to share their experiences in their own words, focusing on events, emotions, and reflections. The interviewer avoids interruptions except when clarification is needed.}

\subsection{Step 2 – Supplementary Questions }
Used only if topics are not covered spontaneously during narration:

\begin{itemize}
        \item \textbf{A. Victim Profile \& Context}
\begin{enumerate}
        \item Had you ever been in an (online) romantic relationship before this experience?
        \item At the time of the incident, how much time did you typically spend online or on social media each day?
        \item For what purposes did you use social media?
        \item What stage of school were you in during that period (e.g., early high school, final year)?
        \item How would you describe your family and social environment (e.g., conservative, traditional, open-minded)? (Encourage the participant to describe this in their own words.)
        \item What made you interested in starting an online relationship at that time?
    \end{enumerate}

    \item \textbf{B. Scammer Profile \& Interaction}
        \begin{enumerate}
        \item How did the person first contact you or start the relationship?
        \item What did they do or say that made you trust them?
        \item In what ways did you later realize the person was not who they claimed to be?
        \item What do you think they were trying to get from you?
    \end{enumerate}
    
    \item \textbf{C. Nature of the Scam}
    \begin{enumerate}
        \item Approximately how long did your online relationship last?
        \item Did any offline meetings or attempts to meet occur?
        \item Was there any exchange of money or gifts? If yes, how much or what kind?
        \item Did the relationship involve any form of sexual conversation or activity (for example, requests for photos or videos)?
    \end{enumerate}
    
    \item \textbf{D. Suspicion and Discovery}
    \begin{enumerate}
        \item At any point, did you start to suspect that something was wrong?
        \item If yes: What made you suspicious? What did you do to find out the truth?
        \item If no: Why do you think you didn’t suspect them earlier? What made them seem trustworthy?
    \end{enumerate}
    
    \item \textbf{E. Aftermath and Recovery}
    \begin{enumerate}
        \item What happened after you discovered the deception?
        \item How did this experience affect your emotions, relationships, and daily life?
        \item What steps did you take to cope or recover (for example, talking to friends, family, or a psychologist)?
        \item Did you seek help from anyone? Why or why not? Who helped you the most, if anyone?
        \item Have your online behaviors or attitudes toward online relationships changed since    
    \end{enumerate}

\end{itemize}

\subsection{Notes for the Interviewer}

    \begin{itemize}
        \item Encourage storytelling rather than short answers.
        \item 
        \item Use gentle follow-up prompts such as:
                \begin{itemize}
                    \item \textit{“Can you tell me more about that?”}
                    \item \textit{“How did that make you feel?”}
                \end{itemize}

    \end{itemize}
    

\section{Appendix: Tables}
\subsection{Victimization Types Reported by Participants \& Participants' Overview}

\begin{table*}
 \caption{Victimization Types Reported by Participants}
  \label{tab:victimization_types} 
  \begin{tabular}{p{4cm} p{3cm} p{5.5cm}}
   \toprule
    \textbf{Victimization Type} & \textbf{Participants} & \textbf{Description} \\
    \midrule
    Emotional damage with or without financial/sexual impact & All & All participants reported emotional abuse. \\
    Financial scam (money, buy gift, re-charge phone sim-card)& Melisa, Tiam, Amin & Participants were scammed out of money. \\
    Sexual Coercion & Ziba, Mahi, Sona, Lina, Mehri, Nina, Sarah, and Hiva & Participants were manipulated into participating in sexual acts (e.g., sexting). \\
    Combination & Sheri, Maria, and Shiva & Participants were manipulated into participating in combined sexual acts and financial assistance. \\
   Only Emotional Victimization & Lili, Zahra & Participants reported emotional victimization -fake profile\cite{whitty2013scammers}. \\
    \bottomrule
  \end{tabular}
\end{table*}

\begin{table*}
  \caption{Participant Overview}
  \label{tab:participant_overview}
  \begin{tabular}{
    p{1cm}
    p{2.5cm}
    p{2cm}
    >{\centering\arraybackslash}p{2cm}
    p{2.5cm}
  }
    \toprule
     P\# & Participant (Gender) & Age (Incident | Interview) & Duration & From \\
    \midrule
    P1 & Melisa (F) & 17 | 19 & 2 months & Urban Area \\
    P2 & Sheri (F)  & 16 | 20 & 4 years  & Urban Area \\
    P3 & Lili (F)   & 17 | 19 & 2 months & Rural Area \\
    P4 & Sona (F)   & 16 | 18 & 5 months & Rural Area \\
    P5 & Lina (F)   & 16 | 19 & 3 years  & Rural Area \\
    P6 & Mehri (F)  & 17 | 18 & 1 year   & Urban Area \\
    P7 & Mahi (F)   & 14 | 18 & 3 months & Rural Area \\
    P8 & Hiva (F)   & 15 | 18 & 6 months & Rural Area \\
    P9 & Nina (F)   & 17 | 25 & 1 year   & Rural Area \\
    P10 & Maria (F) & 16 | 26 & 2 years  & Rural Area \\
    P11 & Sarah (F) & 16 | 19 & 3 months & Rural Area \\
    P12 & Zahra (F) & 13 | 19 & 3 years  & Urban Area \\
    P13 & Ziba (F)  & 16 | 19 & 3 months & Rural Area \\
    P14 & Shiva (F) & 16 | 19 & 3 months & Rural Area \\
    P15 & Amin (M)  & 17 | 19 & 2 months & Urban Area \\
    P16 & Tiam (M)  & 17 | 20 & 6 months & Urban Area \\
    \bottomrule
  \end{tabular}
\end{table*}

\end{document}